\documentstyle[12pt,espcrc1,epsfig]{article}

\begin{document}
\newcommand{\be}{\begin{eqnarray}}
\newcommand{\ee}{\end{eqnarray}}
\begin{center}
   {\large \bf THE QCD PHASE TRANSITION: }\\[2mm]
   {\large \bf FROM THE MICROSCOPIC MECHANISM TO SIGNALS}\\[5mm]
    {\large Talk at Renconres de Moriond - 97 }
   E.V. SHURYAK\\[5mm]
   {\small \it  Department of Physics\\
   SUNY at Stony Brook, Stony Brook, NY11790, USA \\[8mm] }
\end{center}

\begin{abstract}\noindent

  This talk consists of two very different parts: the first one deals with
non-perturbative QCD and physics of the chiral restoration, the second with
rather low-key (and still unfinished) work aiming at obtaining EOS and other
properties of hot/dense hadronic matter from data on heavy ion collisions.
The microscopic mechanism for chiral restoration  phase transition
 is   a transition from randomly placed tunneling
  events (instantons) at low T to a set
of strongly correlated tunneling-anti-tunneling
events (known as instanton - anti-instanton molecules)
 at high T. Many features of the transition 
 can be explained in this simple picture, especially the critical
line and its dependence on quark masses. This scenario  
 predicts qualitative change of the basic quark-quark interactions
 around the phase transition line, with some states (such as pion-sigma ones)
probably surviving even at $T>T_c$.
In the second half of the talk we discuss  experimental
data on collective flow in heavy ion collision,  its hydro-based description
and relation
to equation of state (EOS). A distinct feature  
 of the QCD phase transition region is high degree of ``softness'', 
(small ratio pressure/energy density). We present some preliminary results
indicated that it is indeed needed to explain the radial flow at SPS energies.

\end{abstract}

\newpage
\section{New mechanism for the chiral phase transitions}

  The  QCD-like theories with variable number of colors $N_c$ and (light) 
flavors $N_f$ have
very rich phase structure, which only now starts emerging from
theory and lattice
simulations. It shows
 how naive are many  textbook-style 
explanations, considering  ``overlapping
hadrons'' and ``percolating
quarks''. 
However, in all cases transitions happen at rather
dilute stage of the hadronic phase\footnote{ 
Pure SU(3) gluodynamics
is an especially good example:
 at  $T_c\approx 260 MeV$ the density of glueballs is  negligible
  since even the lightest one has a mass of about 1.7 GeV!}. Naive
geometric ideas cannot explain
 why light fermions play such an important role. In particular,
 the critical temperature $T_c$ for pure gauge simulations
 ($N_f=0$) and those with
dynamical quarks  ($N_f=2-3$) differ
by almost factor 2, being
$T_c\approx 260 MeV$ and  
  $T_c\approx 150 MeV$
 respectively. Furthermore,  at $T\approx T_c$ many  physical
 quantities are
very sensitive to such little details as
 the mass value of the strange quark. (That is why lattice results are
 still
not quite definite about the $order$ of the transition in the real world.)
Furthermore, on the phase diagram as a function of
quark masses there are 2 distinct region of the 1-st order transitions:
the one at large masses is referred to as
   ``deconfinement'' and  the one at low masses ``chiral restoration\footnote{
These different names are partially misleading.
We remind that for light fermions deconfinement has no well defined order parameter: nevertheless in both phase transitions the high-T phase is close to
perturbative quark-gluon plasma, at least this is what EOS data tell us.}''.

  An important
 aspects of the chiral restoration
is
the role of the U(1) chiral
symmetry, see  \cite{u1}. The standard arguments suggest that
the $\sigma$ meson (the
$SU(N_f)_A$ chiral partner of the pion) becomes massless at $T\rightarrow T_c$.
The question is what happens with their  U(1) chiral partners, $\eta'$ and
isovector scalar (with the old name $\delta$ and new one $a_0$) do. 
It remains unknown what happens with thir masses, and some lattice data
suggest those become very light as well.

  Apart of (i) increasing the temperature T, there are other things
one can do to a QCD  vacuum,
and see when (if at all) the quark  condensate disappear and 
 chiral symmetry is restored. 
 For example, one can (ii) increase the baryon density $n_b$, 
(iii) increase the number  of 
quarks $N_f$ in the theory, or (iv) apply the  magnetic field. 
 There are indications that the mechanism I am going to discuss work
in all cases (i-iii). However, in contrast to superconductivity, magnetic field does not
destroy the quark condensate  at all (see \cite{H} and references therein).

   By now
there is large amount of evidence that in vacuum
the 
chiral breaking and other phenomena related with hadrons made of light
quarks are dominated by instantons, see recent review \cite{SS_96}.
 The instanton-based models explain multiple correlation
functions and hadronic spectroscopy, and are also directly supported by
 lattice studies, e.g. \cite{CGHN_94}.

  Now we are ready to discuss the   mechanism underlying chiral restoration.
It is structural
$rearrangement$ of ensemble of
instantons, from relatively random liquid at low T to 
a gas  of  instanton anti-instanton
``molecules''\footnote{Note a similarity to  Kosterlitz-Thouless
  transition in O(2) spin model in 2 dimensions: there are paired
  topological objects (vortices) in one phase and random liquid in
  another.}. 
  In the case of high-T transition one can explain    
 what happens at $T\approx T_c$ rather simply\footnote{  
Unfortunately, in rather technical terms. Simpler explanation is as follows:
tunneling and antitunneling events at $T>T_c$ happen at about the same place.
} 
  Recall that finite T field theory can be described in Euclidean
   space-time by simply imposing finite
periodic box in time, with the Matsubara period $\tau=1/T$. 
A rising T means shrinking box size, and  it turns out the transition
happens exactly when  one  $\bar I I$ molecule fits into
the box. In a series of recent numerical simulations 
of the instanton ensembles (see review \cite{SS_96}) 
this phenomenon is very clearly observed. 
  Not only we get chiral restoration at the right temperature, the
order
of the phase transition and its dependence on quark masses agrees with
lattice data.
 Transition looks like the $second$ order  for
$N_f=2$ massless flavors, but turns to a {\it weak first order} one for QCD
with physical masses.
 Furthermore, the thermodynamic parameters, the
spectra of the Dirac operator, the T-dependence
 of the quark condensate and various susceptibilities, 
the screening masses  and even the critical
 indices are 
consistent with available lattice data.

\begin{figure}[t]
\begin{center}
\includegraphics[width=12.cm]{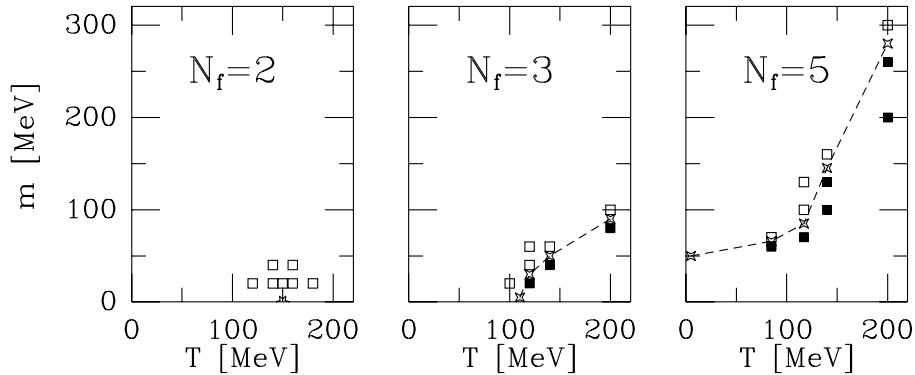}
\end{center}
\caption{\label{myphases}
Phase diagram (temperature-quark mass plane) of the instanton liquid for
different numbers of quark flavors, $N_f$=2,3 and 5. 
The open squares indicate
 non-zero
chiral condensate, while  solid one indicate that it
is zero. The dashed lines 
show the approximate location of the discontinuity line.
}\end{figure}

  Few words about 
QCD
with larger $N_f$. Fig.\ref{myphases} show that
 at  $N_f=5$
instantons can no longer  break
the chiral symmetry\footnote{Note
that the $N_f=4$ case is  missing, because the
condensate is small and comparable to finite-size effects. Amusingly, recent
results from Columbia group for
$N_f=4$ have found exactly this:
  a dramatic  drop in chiral symmetry
 breaking effects (e.g. $\pi-\sigma,\rho-a_1,N-N^*(1/2^-)$
 splittings.}  even at T=0  (provided quarks are light
 enough).
. On the other hand, if instantons cannot break chiral
symmetry, other effects like confinement or even
one gluon exchange can do it, and according to
 \cite{ATW_96} 
this happens for $N_f\simeq 11$. Existing estimates
\cite{coulombcond} show that the
quark condensate in the domain $N_f=5 - 11$ is small, with a value from 
  about 1/10 of the QCD value to the exponentially vanishing one a t
the upper end. Lattice measurements in this domain so far does not see
it, and  report chirally symmetric ground state at 8,12 and 16 flavors.

  The last topic is hadron modification near the phase transition line.
Their masses are expected to change as
quark condensate decreases, and 
 the most radical idea is the Brown-Rho scaling, suggesting that
 $all$ hadronic masses get their scale from the quark condensate,
 and therefore vanish  
at $T\rightarrow T_c$.
However
in a vacuum containing instanton molecules it is not so, because these objects
generate new non-perturbative
inter-quark interactions. The
 Lagrangian  describing those is discussed in \cite{SS_96}: in some channels like $\pi,\rho$ it leads to  attractive forces which (if strong enough)
may creat bound states even $above$ the transition line.
So far it is unclear whether it happens, but existence of these forces can be checked using the 
 so called ``screening masses''. Their 
T-dependence for a number of hadronic channels
 was calculated in the interacting instanton model, and those
 show good  agreement with lattice
ones. Especially important is strong attraction in scalar-pseudoscalar
channels, shifting these masses down from their high-T asymptotic,
$M/\pi T=2$.

\section{Looking for Equation of state in high energy  heavy  ion collisions }

  Recent data obtained with heavy ion beams (Au at Brookhaven and Pb at 
CERN) have displayed very strong collective flow effects (see other talks and
my summary). These data can be used in order
to obtain information
  about properties of the Equation of State
(EOS) of hot/dense hadronic matter.

  In order to do so, it is useful to
return back from a (very complicated)
 cascade event generators  to basic
hydro description\footnote{Hydro does not contradict
to event generators, but rather get support from them. It just allows to
get space-time picture and flow much easier, without simulating
all multiple re-scattering in more-or-less thermal conditions.}
and can easily incorporate different scenarios (e.g., with or without
the QCD phase transition). 
In contrast to longitudinal flow, for $radial$ direction
  we know the
  initial conditions rather well, and therefore quantitative relation 
between the  magnitude of the radial flow in central collisions and EOS
can be made.

\begin{figure}[h]
\includegraphics[width=8cm,angle=-90]{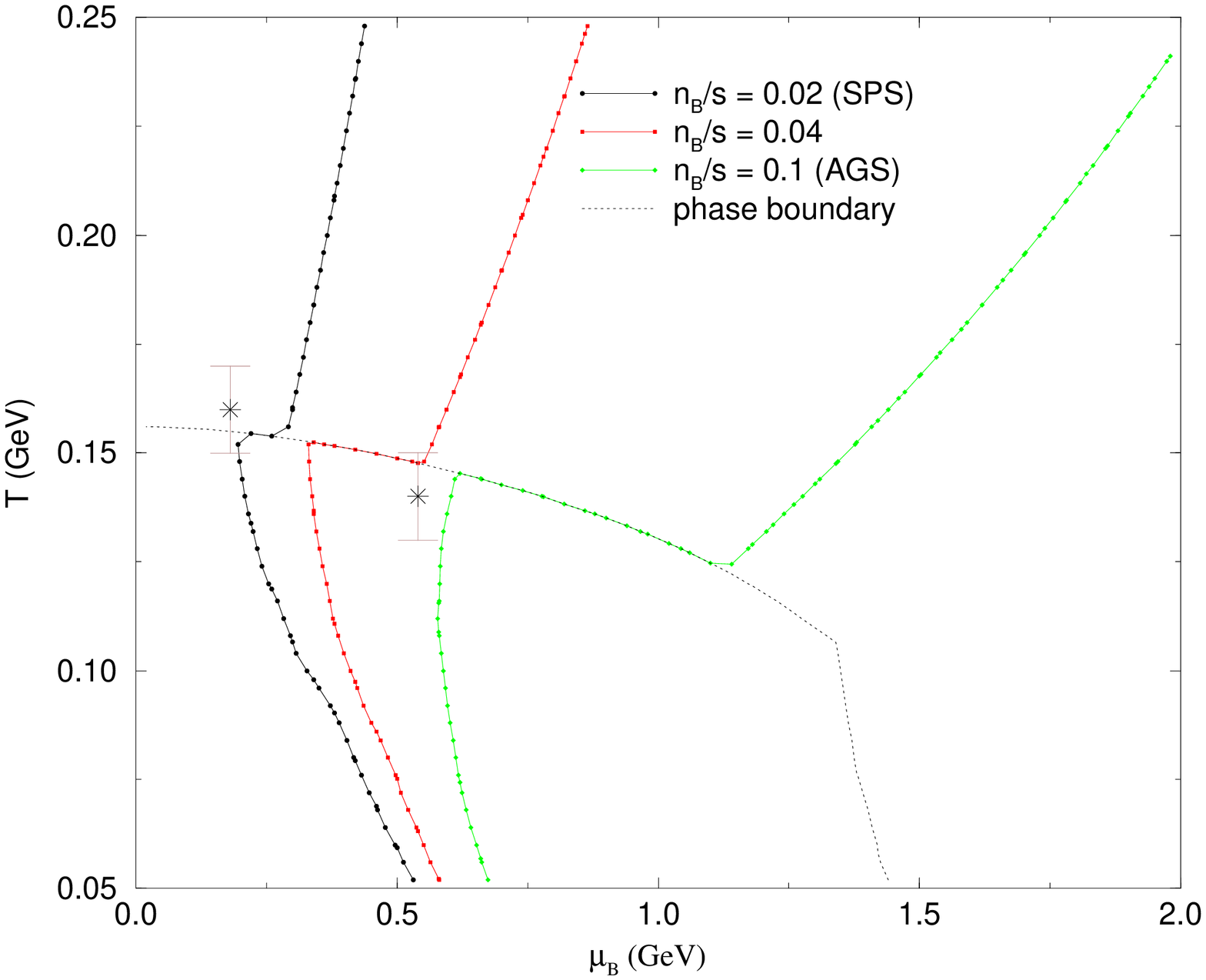}
\includegraphics[width=8cm,angle=-90]{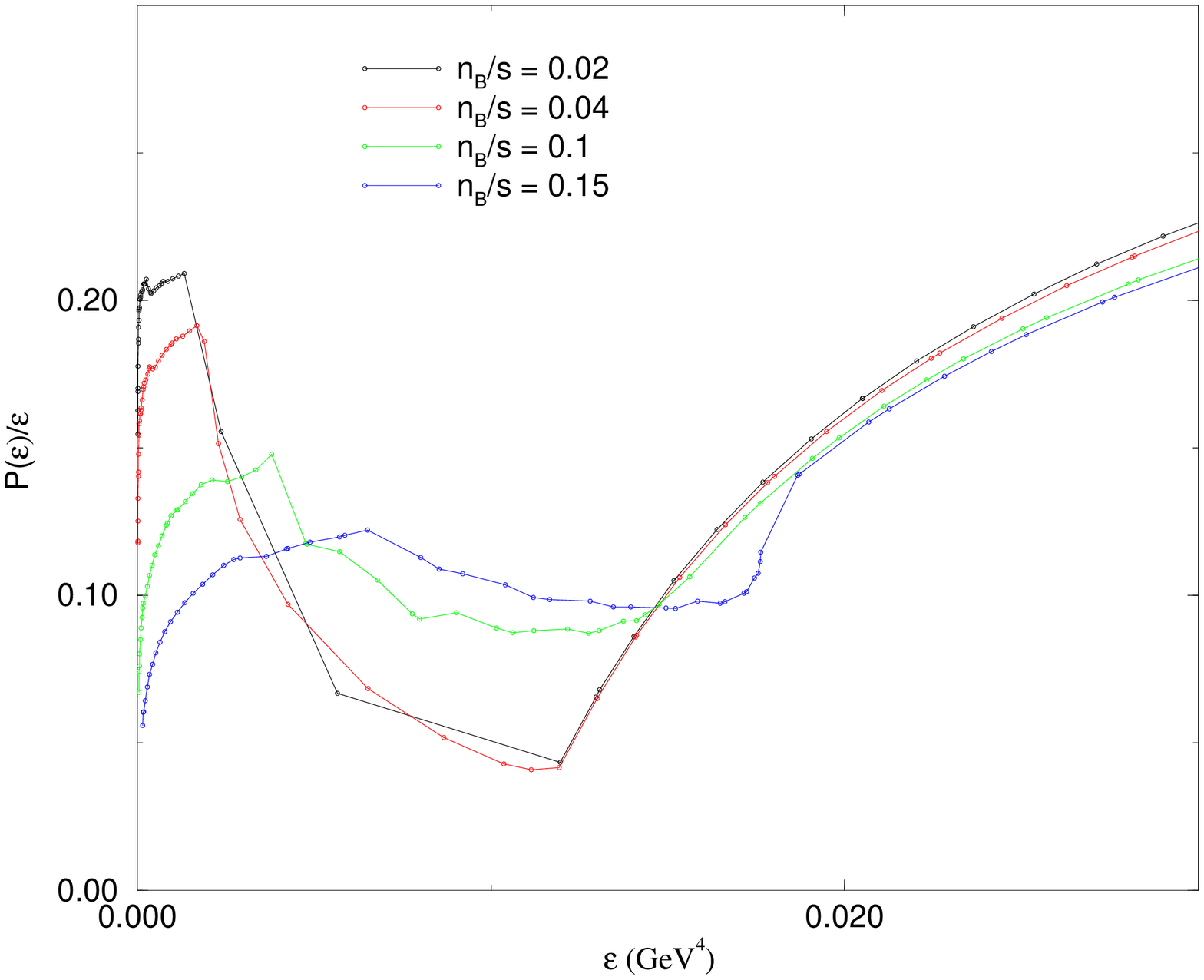}
\caption{\label{fig_EOS}
 (a) Paths in the $T-\mu$ plane for different
baryon admixture, for resonance gas plus the QGP; 
(b) the ratio of pressure to energy density $p/\epsilon$ versus
$\epsilon$,
for different baryon admixture. }
\end{figure}

   Let me show some results from (still unfinished) work by M.Hung and myself,
which is aiming at developing
 a new model for AGS/SPS energy domain, called {\it Hydro-Kinetic
   Model}, HKM. 
   The  basic Equation of State (EOS) of hadronic matter 
used is that of a {\it resonance gas}, while for QGP one usually uses 
a simple bag-type EOS, with a constant fitted to $T_c= 160 MeV$.
  	In Fig.\ref{fig_EOS}(a) we show that phase boundary and the
 paths on the phase
diagram. As both baryon number and entropy is conserved, the lines are
marked by their ratio. Those for
$n_b/s=0.02,0.1$ correspond approximately to SPS (160 GeV A) and AGS (11 GeV A) heavy ion
collisions,
respectively. Note that the trajectory has a non-trivial zigzag shape,
with re-heating in the mixed phase.
The endpoint of the QGP branch is known as the ``softest
point'' \cite{HS_95}, while the beginning of the hadronic one we will call the
``hottest
point''\footnote{Of course, in the ``Hagedorn sense'', as the hottest
  point of the hadronic phase.}.

Experimentally observable particle composition is related to the
stage of the collision  known as a $chemical$
freeze-out.  
 Multiple works (e.g. \cite{PBM_etal}) have applied thermal description and
 determined where those points are on the $T,\mu_b$  phase diagram:
for both AGS and SPS those (inside error bars) coincide with the ``hottest
points" on our zigzag.

  The next step is to define the effective EOS
in the form $p(\epsilon)$ (needed for hydro) {\it on these lines}:
that is shown in Fig.\ref{fig_EOS}(b)
.
Note that  the QCD resonance gas 
in fact  has  a very simple EOS
$p/ \epsilon\approx const$, while the ``mixed phase" is very soft indeed.
The contrast between ``softness" of matter at dense stages and relative 
``stiffness"  at the dilute ones is strongly enhanced for the SPS case. 

  The final velocity of the observed collective ``flow''
is time integral of the acceleration, which is proportional to $p/ \epsilon$
ratio plotted in Fig.\ref{fig_EOS}(b). Although the observed velocity is
not very different for both energies, this is kind of a coincidence since both
the EOS and the time development of the collisions are rather different.
In short, it was found that this basic EOS is too soft for AGS data on flow\footnote{It was also known from 
RQMD-based studies that extra repulsive interaction between baryons is
indeed need to reproduce radial flow.},
but describe well the SPS case. Most interesting, if one assume that
there is $no$ phase transition at all, and $p/\epsilon\approx .2$ like in the resonance gas, the flow obtained is way too strong.

    The most difficult puzzle related to radial flow is
 provided by experimental data showing
that it has very strong
  A-dependence (see e.g. talk by Gaardhoje in this proceedings):
 the larger the nuclei, the stronger is the flow.
  To resolve this (and other) puzzle one should correctly include the
kinetics of the freeze-out. In most hydro papers, expansion was cut off
at fixed T, usually about 140 MeV. In other words, it was assumed that
the hadronic matter dries out very quickly, and its large pressure
cannot lead to any significant motion. The situation is different if one  
apply correct kinetic condition:
 the 
(relevant) collision rate approach 
 the expansion rate. 
Then one finds that
 with new high energy heavy ion
beams (Pb at CERN and Au at BNL) we now have access to $cooler$
hadronic gas, compared to medium ion
collisions studied few years ago. The temperature of $thermal$ freeze-out
id only slightly smaller, 110-120 MeV, but in terms of space-time evolution
this difference imply a very significant change. For sufficiently heavy
ion collisions the matter  gets the ``extra push" from stiff pion gas at the end, which explains large flow velocity.

 One interesting effect\footnote{It was originally pointed
out  by G.Baym,
see details in   \cite{Bebie_etal}.}
 which follows from this observation: at the expansion 
  stage without chemical but  in
   thermal equilibrium the chemical potentials of all species change.
The 
 chemical potentials of the pions\footnote{For
  clarity: those potentials are conjugated to  
total number of particles, so say for pions they enter distributions of
$\pi+,\pi^-,\pi^0$  with the same sign. }
 created in  central PbPb  it should reach
$\mu_\pi=60-80 MeV$. NA44 data indeed support
extra low-$p_t$ enhancement in PbPb relative to S-induced
reaction.

  All phenomena discussed are predicted to be 
very much enhanced at RHIC. Larger multiplicity lead to (paradoxically) smaller final temperatures, and softness of the EOS due to the QCD phase transition change the  lifetime of matter by a factor
of 2 \cite{RG_96}. 

\section*{Acknowledgments}
This talk is based on works done
my collaborators,
 especially with T.Schaefer and C.M.Hung.
This work is partly supported by the 
US Department
 of Energy under Grant No. DE-FG02-88ER40388.


\end{document}